\documentclass[10pt,conference]{IEEEtran}  
\IEEEoverridecommandlockouts

\usepackage[english]{babel} 
\usepackage{caption}
\usepackage{amsmath}
\usepackage{subfigure}
\usepackage{graphicx}
\usepackage{multirow}
\usepackage{algpseudocode}
\usepackage{algorithm}
\usepackage{amssymb}
\usepackage{balance}
\usepackage{color}
\usepackage{multirow}
\usepackage{multicol}
\usepackage{listings}
\usepackage{subfigure}
\usepackage{float}
\usepackage{enumerate}
\usepackage{cite}  
\usepackage{wasysym}
\usepackage{hyperref}
\usepackage{booktabs}
\usepackage{enumitem} 
\usepackage{balance}
\usepackage[all]{nowidow}

\usepackage{colortbl}
\definecolor{mygray}{gray}{0.92}

\renewcommand{\baselinestretch}{0.97}

\newcommand{\mytool}{{VPDroid}}

\usepackage{tikz}
\newcommand*\circled[1]{\tikz[baseline=(char.base)]{
            \node[shape=circle,fill,inner sep=1pt] (char) {\textcolor{white}{#1}};}}

\usepackage{xcolor}

\definecolor{mGreen}{rgb}{0,0.6,0}
\definecolor{mGray}{rgb}{0.5,0.5,0.5}
\definecolor{mPurple}{rgb}{0.58,0,0.82}
\definecolor{backgroundColour}{rgb}{0.95,0.95,0.92}

\lstdefinestyle{CStyle}{
    backgroundcolor=\color{backgroundColour},
    commentstyle=\color{mGreen},
    keywordstyle=\color{magenta},
    stringstyle=\color{mPurple},
    basicstyle=\scriptsize\ttfamily,
    breaklines=true,
    captionpos=b,
    keepspaces=true,
    otherkeywords={status_t, ArrayObject,uint32_t},
    numbers=left,
    numbersep=1pt,
    showspaces=false,
    showstringspaces=false,
    showtabs=false,
    tabsize=2,
    language=C
}

\begin{document}

\title{App's Auto-Login Function Security Testing via Android OS-Level Virtualization\\
{\footnotesize}
\thanks{{*}Corresponding author: guojpeng@whu.edu.cn.}
\thanks{\dag(1) Key Laboratory of Aerospace Information Security and Trust Computing, Ministry of Education;}
\thanks{(2) School of Cyber Science and Engineering, Wuhan University.}
}

\author{\IEEEauthorblockN{Wenna Song\dag}
\IEEEauthorblockA{\textit{Wuhan University} \\
Wuhan, China \\
swenae@whu.edu.cn}
\\
\IEEEauthorblockN{Yi Xiang}
\IEEEauthorblockA{\textit{Wuhan University} \\
Wuhan, China \\
xiangyi@whu.edu.cn}
\and
\IEEEauthorblockN{Jiang Ming}
\IEEEauthorblockA{\textit{The University of Texas at Arlington} \\
Arlington, TX, USA \\
jiang.ming@uta.edu}
\\
\IEEEauthorblockN{Yuan Chen}
\IEEEauthorblockA{\textit{Wuhan University} \\
Wuhan, China \\
sairen@whu.edu.cn}
\and
\IEEEauthorblockN{Lin Jiang}
\IEEEauthorblockA{\textit{Independent Researcher} \\
Xian, China\\
pppaass@163.com}
\\
\IEEEauthorblockN{Jianming Fu\dag }
\IEEEauthorblockA{\textit{Wuhan University} \\
Wuhan, China \\
jmfu@whu.edu.cn}

\and
\IEEEauthorblockN{Han Yan\dag}
\IEEEauthorblockA{\textit{Wuhan University} \\
Wuhan, China \\
cool.yim@whu.edu.cn}
\\
\IEEEauthorblockN{Guojun Peng{*}\dag}
\IEEEauthorblockA{\textit{Wuhan University} \\
Wuhan, China \\
guojpeng@whu.edu.cn}

}
\maketitle


\begin{abstract}

Limited by the small keyboard, most mobile apps support the automatic login feature for better user experience.
Therefore, users avoid the inconvenience of retyping their ID and password when an app runs
in the foreground again. However, this auto-login function can be exploited to launch
the so-called ``data-clone attack'': once the locally-stored, auto-login depended data are cloned by attackers and placed into their own smartphones,
attackers can break through the login-device number limit and log in to the victim's account stealthily.
A natural countermeasure is to check the consistency of device-specific attributes.  As long as the new device
shows different device fingerprints with the previous one, the app will disable the auto-login function and thus prevent data-clone attacks.

In this paper, we develop \emph{VPDroid}, a transparent Android OS-level virtualization platform tailored for security testing.
With VPDroid, security analysts can customize different device artifacts, such as CPU model, Android ID, and phone number,
in a virtual phone without user-level API hooking. VPDroid's isolation mechanism ensures that user-mode apps in the virtual phone
cannot detect device-specific discrepancies. To assess Android apps' susceptibility to the data-clone attack,
we use VPDroid to simulate data-clone attacks with $234$ most-downloaded apps.
Our experiments on five different virtual phone environments show that VPDroid's device attribute customization
can deceive all tested apps that perform device-consistency checks, such as Twitter, WeChat, and PayPal.
19 vendors have confirmed our report as a zero-day vulnerability.
Our findings paint a cautionary tale: only enforcing a device-consistency check at client side is still vulnerable to an advanced data-clone attack.

\end{abstract}


\section{Introduction} \label{sec:intro}

With the prosperous development of the Android system and mobile networks~\cite{AOS,Mnet},
the apps running on Android keep updating constantly to meet the
fast-growing demand of smartphone users. In addition to the standard functionalities
such as communication and entertainment, apps are now performing various critical tasks
such as social networking~\cite{Omlet14}, GPS navigation~\cite{Canino18}, IoT device remote control~\cite{Ruiz18},
and mobile payment~\cite{Sherman14}. Inevitably large amounts of private data (e.g., user credentials) are stored in the smartphone.
Therefore, 
the cyber arms race between bypassing user authentication and its countermeasure
has transformed into an intensive tug-of-war.


According to Verizon's 2019 data breach investigation report~\cite{verizon19},
``76\% of network intrusions exploited weak or stolen credentials.''
Over the past decade, the attacks to take over smartphone user accounts
also generated a large body of literature.  We particularly examine the high-impact attacks and find out that,
their root causes lie in either fundamental design flaws or the system's underlying vulnerabilities.
Just as severe security vulnerabilities in Android password manager apps allow attackers to access the stored credentials~\cite{Huber17A,OAuthLint}, man-in-the-middle (MitM) attacks exploit the password reset vulnerability to crack a mobile user's account password~\cite{gelernter17,WangDong18},
and the recently developed app-virtualization technique defies Android unique user ID mechanism, causing guest apps vulnerable to
the ``shared-everything threat''~\cite{DualInstance,Middle,Deshun20}.
In this paper, we focus on the security risk caused by mobile apps' auto-login functions,
which belongs to client-side tampering vulnerabilities~\cite{LukKim20}.

Most of the existing mobile apps support automatic login to optimize the user experience.
It avoids the hassle of retyping user ID and password in a small keyboard when reaccessing the app.
Mobile users have gotten used to using the auto-login feature due to its convenience.
However, if an attacker steals the auto-login depended data from a user's device
and replaces his data with the user's, the attacker can bypass
the login-device number limit and access the user's account without raising suspicion---we call it as a ``data-clone attack.''
In this paper, the meaning of "credential" or the auto-login depended data is a token or user identity.
Initial investigations have studied this threat~\cite{Junsung18,Jongwon15,Suwan16}, but all of them
are limited to victim identity theft on a rooted device. Furthermore, they missed an important fact:
an increasing number of apps check device consistency to prevent client-side tampering; if they detect any device-specific discrepancies,
their auto-login functions will be disabled. 

We present a new attack model that can break through the paying-subscriber limit on non-rooted devices.
For many subscription-based apps~\cite{subscription-app}, such as Netflix, Amazon Prime Video, and Apple Music,
their revenue models impose a maximum number of the same user's login from different devices at a time.
For example, Netflix's Basic plan only allows to stream high-definition (HD) video
on one device at a time. A fraudster first pays Netflix's Basic plan fee.
Then, he leverages an OEM-made phone clone app~\cite{phone-clone-app1}
to launch a data-clone attack. The OEM-made phone clone app can copy private data
between the same OEM phones without rooting devices.
In this way, the fraudster enjoys
the Premium plan service---watching HD video on multiple screens at the same time.
This new attack model can even enable non-paying customers to use premium services completely free of charge.

To assess Android apps' susceptibility to data clone attacks, we perform an empirical study on
$234$ most-downloaded apps from American and Chinese Android app markets (\textbf{Study 1}). Study 1 is presented in III.C: we perform data-clone attacks with real devices.
Our tested apps have billions of users in total.
After performing data clone attacks, we can successfully bypass user authentication and access $131$ apps, including Facebook, Snapchat, QQ, and Weibo.
We further study the failure causes of the remaining $103$ apps and find out that they have already taken actions to
secure the auto-login function. The most common strategy is to check the consistency of device footprints when the app is resumed,
such as checking Android ID, MAC address, and International Mobile Equipment Identity (IMEI).
If any device-specific discrepancies are detected, the app will disable the automatic login,
and users have to retype their ID and password manually.
Some critical apps (e.g., PayPal) can even fingerprint a rooted device and the Android runtime hooking framework, Xposed~\cite{Xposed},
which can create an app-virtualization environment to modify device attributes.

Not wanting to stop there, we explore Android OS-level virtualization to
develop a transparent device-attribute editing platform, named \emph{\mytool}.\footnote{``VPDroid'' means running a Virtual Phone on Android system.}
\mytool\ provides a customizable, native-performance virtual phone (VP)
environment on a single physical device. The VP runs a standard Android environment,
and security analysts can configure the VP with up to $101$ options to simulate a smartphone's profiles.
\mytool\ facilitates testing an app's capability of detecting the change of device, as well as the resilience against
the data-clone attack.

\mytool\ is a heavily modified version of Cells~\cite{Cells}, which is the first mobile OS virtualization solution to
support running virtual phones on a single OS instance.
Unfortunately, Cells's virtualization methods to many hardware devices (e.g., filesystem, network, display, and power)
have been obsolete since Android 6.0. Besides, Cells lacks virtualization support for Bluetooth, GPS, and Android Debug Bridge (ADB).
We improve the multiplexing of hardware devices in two ways: 1) rewriting kernel drivers (e.g., GPS) to adapt to new Android version updates;
2) we develop a new user-level device virtualization mechanism to virtualize proprietary devices, which are completely closed source.
More importantly, \mytool\ supports editing the VP's device-specific attributes.
We carefully design where to edit device attributes---all customization functions are
executed outside of the VP.
Our isolation design ensures that a user-mode app running in the VP is unaware of device-specific differences.

\mytool\ has been tested to work seamlessly across Android 6.0 and Android 10.0. We repeat Study 1 in Section VII but on top of the custom virtual phone.
We install \mytool\ in a Google Nexus 6P smartphone and redo \textbf{Study 1} by configuring the VP as
five different environments: Xiaomi Redmi Note 4, Redmi Note 4x, Huawei Honor 6x, Honor 8, and Google Nexus 6P.\footnote{Two Nexus 6P phones are different in CPU model and ROM size.}
In each VP, we can compromise all of the $234$ most popular apps' accounts, including $103$ apps that perform device-consistency checks.
We have made responsible disclosure to the app vendors, and 19 of them have confirmed our report
as a zero-day vulnerability. At last, we discuss possible countermeasures to defeat data-clone attacks.
Our study demonstrates that only enforcing device-consistency checks is still vulnerable to an advanced data-clone attack.
In a nutshell, we make the following three significant contributions:


\begin{itemize}
	\item  Our work reveals the security risk of Android apps' auto-login
           functions. In a addition, we introduce a new attack model that can break through the paying-subscriber limit on non-rooted devices.
    \item  We improve the Android OS-level virtualization technique to develop a transparent device-attribute editing platform.
           Security analysts can simulate more diversified VPs on a single device. All of our tested apps are deceived into thinking that the device is not changed.
	\item  \mytool\ has broad applications that rely on a virtual phone environment. Our clone attack demo video (\url{https://youtu.be/cs6LxbDGPXU})
           shows that \mytool\ enables the attacker to bypass KakaoTalk's device-consistency check, and the victim is unaware that her account has been compromised.
           We release \mytool's source code at (\url{https://github.com/VPDroid/Dev}).
\end{itemize}



\section{Background and Related Work}
\label{sec:background}


In this section, we first discuss the security risk of automatic login in mobile apps.
Existing works on exploiting Android apps' auto-login functions are limited.
Then, we introduce OEM-made phone clone apps, which we take as a vector to clone private data.
At last, we describe the principle of Android OS-level virtualization,
which is the foundation of \mytool.

\subsection{Automatic Login in Mobile Apps}

Limited by the small-scale touchscreen, typically, only one app is running in the foreground of a smartphone, and users frequently switch to
other apps in the background. It is rather cumbersome having to type ID and password every time users access an app.
To optimize the user experience, most mobile apps support the automatic login feature by default. As a result, users only need to input
their ID and password at their first login time. After that, users can access the app smoothly without retyping their ID and password.
For all of our tested $234$ apps, their auto-login functions still work even when we kill their processes and restart them later.

Most auto-login functions store user credential data locally and complete the authentication process with the app server automatically.
User credential data are the security tokens used to certify a user's identity with the app server.
After the user first enters ID and password to go through the authentication process, the app locally stores
user credential data for future verification purposes.
Android provides four options to save app-private data~\cite{android-storage}:
1) internal file storage, 2) external file storage, 3) shared preferences, and 4) databases. User credentials are typically stored either
in the form of key-value pairs as \texttt{SharedPreferences} or structured data in an SQLite database.  Both of them are under the private directory of
``/data/data/[app\_name]/'', and other apps typically do not have the privilege to access them. 

\subsection{Exploiting Auto-login Function Works and Limitations} \label{sec:related-limitation}


If attackers steal the locally-stored, auto-login depended data and put them under the same directory of a different phone,
attackers can leverage the auto-login feature to bypass the authentication from the server side.
This means attacks can automatically log in to the victims' accounts without knowing their ID and password.

We take WeChat, a social media app with over 1 billion daily active users~\cite{wechat-1-billion},
as a case study. WeChat stores AES-encrypted user credentials in a SQLite database file ``EnMicroMsg.db''. This file is under the directory of ``/data/data/MicroMsg/[xxxx...xxxx]/'',
in which ``xxxx...xxxx'' is the 32-bit md5 value of a file name.
In addition to ``EnMicroMsg.db'', we find WeChat's auto-login function also relies on multiple files under the same directory
and a system configuration file, ``/data/data/MicroMsg/systemInfo.cfg''. ``systemInfo.cfg'' is an XML plaintext containing the connection information
with the app server. Apparently, only cloning ``EnMicroMsg.db'' is not enough at all. Note that the exact files that are needed by the auto-login function vary on a case-by-case basis.
Therefore, the best strategy is to clone all of the data under ``/data/data/[app\_name]/'' to the target device.

Recent papers have exploited the pervasive auto-login feature in Android apps~\cite{Junsung18,Jongwon15,Suwan16}.
These studies share two common assumptions:
\begin{enumerate}
  \item The victim's device has been rooted.
  \item Attackers either have physical access to the victim's rooted device, or the malware to steal credential data has been installed on the rooted device.
\end{enumerate}
Rooted Android devices are very common in countries outside of North America, especially in Asian countries.
Tencent research shows that 80\% of Chinese users had a rooted device~\cite{NuData17}.
Besides, according to the official Android security report~\cite{android-2018-report},
large families of harmful applications use privilege escalation exploits to root devices.
These papers~\cite{Junsung18,Jongwon15,Suwan16} demonstrated the feasibility of data-clone attacks with a very small number of apps---only six apps in total.
However, none of them take device-consistency checks into consideration.

We still take WeChat as an example to explain the limitation of existing work~\cite{Junsung18,Jongwon15,Suwan16}.
We clone all of the data under ``/data/data/MicroMsg/'' to a new smartphone, but we still cannot automatically log in to WeChat.
WeChat pops up the login interface and asks us to retype ID and password.
The root cause is the change of a smartphone environment is almost instantly detected by
WeChat, and then it terminates the auto-login process.
We find WeChat detects 22 device footprints such as phone number, IMEI, and Bluetooth address.
In our dataset,  a total of $103$ apps such as Chrome, Apple Music, KakaoTalk, and PayPal also conduct a similar detection
when invoking their auto-login functions.


Bianchi et at.~\cite{Bianchi17} also simulate device-public information to bypass user authentication.
They exploit an entire class of apps that only rely on device-public information to authenticate the user to their backends.
However, in our tested 234 most-downloaded apps, no one adopts such a weak authentication scheme, including WhatsApp and
Viber that were once vulnerable in this paper. Another major difference is that they customize only 13 device-public profiles
in the Xposed framework~\cite{Xposed} by hooking APIs. By contrast, we improve OS-level virtualization to deliver an open-source,
more transparent device-attribute editing platform, which can edit 101 device artifacts without user-level API hooking.

\begin{figure} [t]
\centering
\includegraphics[width=0.43\textwidth]{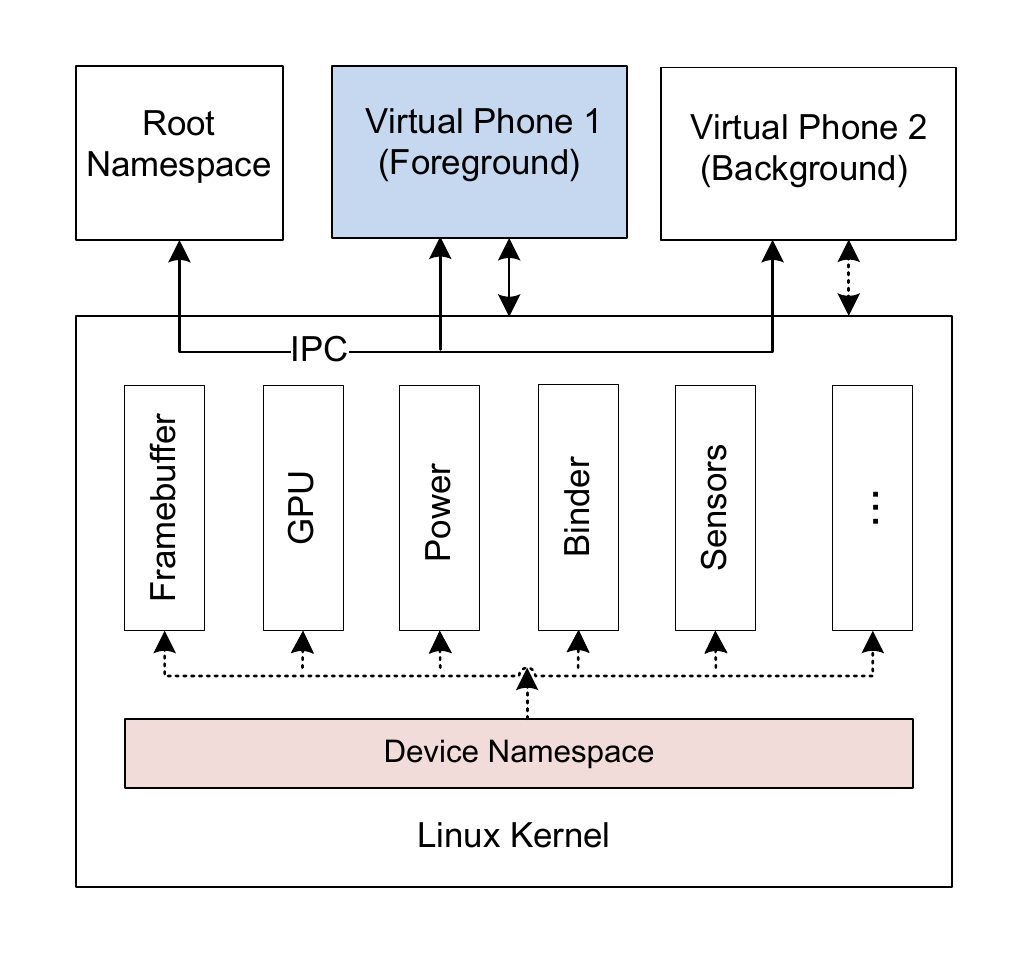}
\caption{Cells kernel-level device virtualization overview. The VP running in the foreground is displayed at any time and always given direct access to hardware devices.}
\label{fig:cells}
\vspace{-5mm}
\end{figure}

\subsection{OEM-Made Phone Clone Apps}


$\S$\ref{sec:subscriber-fraud} will describe paying-subscriber fraud,
in which a fraudster leverages OEM-made phone clone apps to launch
data clone attacks on non-rooted devices.
Phone clone apps are getting popular in various Android app markets.
All of the top Android phone manufacturers~\cite{top-andriod-phone}
custom their own phone clone apps, such as Samsung Smart Switch, Huawei Phone Clone, and Xiaomi Mi Mover.
Most of them have been downloaded more than 100 million times.
These OEM-made phone clone apps have a unique advantage: they have the privilege to call Android Backup API~\cite{android-backup} on the same OEM phones.
Therefore, in addition to contacts, call logs, photos, and data in external storage, they are able to transfer app-private data in ``/data'' partition between the same OEM phones without rooting devices.
For example, Smart Switch can seamlessly transfer app private data and home layouts between Galaxy devices, and it is similar for other OEM-made clone apps.
This advantage brings users great convenience when they upgrade their devices: the cloned smartphone just becomes the replica of the old device,
and apps behave exactly as if they are still on the old device.



\subsection{Android OS Virtualization}

\begin{figure*} [t]
\centering
\includegraphics[width=0.81\textwidth]{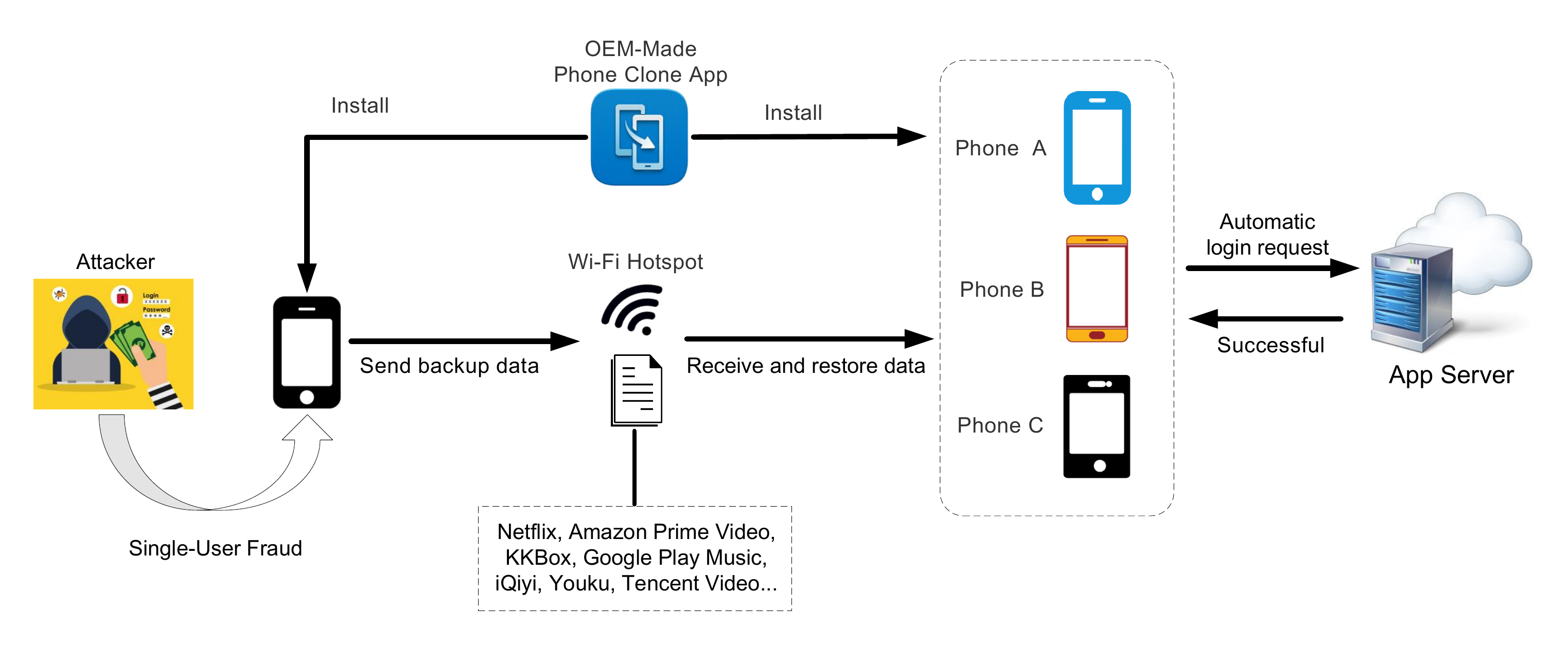}
\vspace{-1mm}
\caption{Data-clone attacks distribute auto-login depended data for the paying-subscriber fraud.}
\label{fig:subscription-based}
\vspace{-6mm}
\end{figure*}

We apply Android OS-level virtualization to editing device-specific artifacts.
Mobile virtualization means running multiple separate instances of smartphone environments on the same
physical device.
Unlike desktop and server machines, resource-constrained mobile devices limit the adoption
of hypervisor-based virtualization~\cite{Barr2010,Dall2014}, while OS-level virtualization~\cite{Soltesz07,Felter15}
becomes an acceptable option. Cells~\cite{Cells} is the first lightweight OS-level virtualization solution
to run multiple isolated virtual phones on a single Android instance. In each virtual phone (VP),
a user can execute unmodified apps and perform normal smartphone operations.
Cells made most hardware device virtualization in the Linux kernel layer, and Figure~\ref{fig:cells} shows an overview of
Cells's kernel-level virtualization architecture.
The VP running in the foreground is displayed at any time and \emph{always given direct access to hardware devices}.
Cells invents a new \texttt{device} namespace mechanism to support efficient hardware resource multiplexing,
and each VP is associated with a unique \texttt{device} namespace for device interaction.
To make various hardware devices aware of \texttt{device} namespaces, Cells virtualizes kernel interfaces in three ways:
1) create a device driver wrapper; 2) modify a device subsystem; 3) modify a device driver.

Unfortunately, Cells's design lacks flexibility. Many heavily-modified kernel drivers are
susceptible to new Android version updates. Since Android 6.0,
Cells's virtualization to many hardware devices has been obsoleted. In addition,
it also lacks device virtualization solutions for Bluetooth, GPS, and ADB;
their artifacts are commonly used to fingerprint different devices.
Even Cells's commercial version, Cellrox\footnote{Cellrox - Mobile Virtualization Platform: \url{http://www.cellrox.com}.},
is only compatible with Android 5.1. Our work bridges the gap in mainstream Android versions.

\section{Paying-Subscriber Fraud \& Device-Consistency Check} \label{sec:data-clone}

In this section, we first describe the unique benefit of exploiting auto-login functions to bypass user authentication.
Next, we introduce a new data-clone attack model: paying-subscriber fraud.
Then, we perform data-clone attacks with 234 most-downloaded apps. Our results show that data-clone attacks are
a real threat to both the app economy and user privacy, especially when skilled attackers are able to simulate a Hi-Fi
smartphone environment.

%

\subsection{Data-Clone Attack's Advantage}

\begin{figure}
\centering
\includegraphics[width=0.5\textwidth]{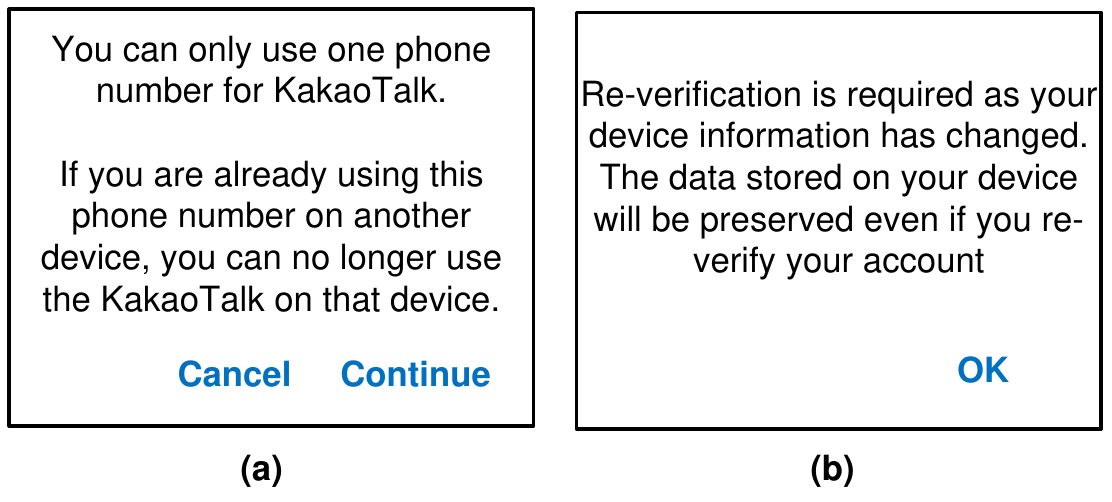}
\caption{KakaoTalk's warning notifications.}
\label{fig:warning}
\vspace{-5mm}
\end{figure}

Compared to the case that the attacker can intercept the user's password,
the unique benefit of the data-clone attack is much stealthier.
The reason is that the login process by typing the user's ID and password would
trigger the detection of the login-device number limit on the server side.
Many apps only allow a single user to log in from one device at a time.
This means the legal user and the attacker cannot be online simultaneously by typing the user's ID and password.
For example, when an attacker logs in to a messaging app, KakaoTalk, from a different phone by typing the victim's ID and password,
the attacker's phone will receive a warning notification as shown in Figure~\ref{fig:warning}(a).
However, our key observation is \emph{counting the number of login devices is not affected by multiple auto-login attempts from the same device},
which leaves a backdoor for us to break through the login-device number limit.
As a result, the user's sensitive data will be in jeopardy without raising suspicion.
If a social messaging app is compromised in this way, the adversary can not only review chat history in real time
but also impersonate the victim to send messages.

Our demo video (\url{https://youtu.be/cs6LxbDGPXU}) shows such an identity theft example.
When the legal user is online, the attacker cannot log in to KakaoTalk by
typing the same user's ID and password.
Furthermore, KakaoTalk can detect the change of a new device. It disables the auto-login
after we copy the data in the directory of ``/data/data/KakaoTalk/''
to a new device (see Figure~\ref{fig:warning}(b)). In contrast, we perform a data-clone attack after we customize \mytool's VP
with the old phone's profiles. We find the victim and the attacker can be online at the same time without raising suspicion.

\subsection{Paying-Subscriber Fraud} \label{sec:subscriber-fraud}

The subscription-based app economy thrives in mobile markets, and customers have acclimated to
the idea of regular payments for a better service~\cite{subscription-app}.
Typical examples are the apps that provide video and music streaming services, such as
YouTube, Netflix, Amazon Prime Video, iQiyi, and Youku Video. For a
subscription-based app, only a paying subscriber can enjoy its premium service,
and it also enforces the maximum number of the same user's login from different devices at a time.
For example, Netflix's premium plan allows at most four screens that a user can
watch on simultaneously.

As counting the number of login devices is typically not affected by multiple auto-login attempts from the same device,
even with a non-rooted device, a fraudster can use an OEM-made phone clone app to perform data-clone attacks
and break through the paying-subscriber limit. Figure~\ref{fig:subscription-based} illustrates such an example,
and eventually, the fraudster can access Android premium apps in multiple devices without payments.
Although Figure~\ref{fig:subscription-based} shows
a single-user fraud case, once this attack model is turning into full-fledged, coordinated attacks in the Android black market,
malicious actors can infringe the revenue model of subscription-based apps, resulting in tremendous financial losses to software vendors.
Most of the zero-day vulnerabilities that we found belong to this category, and
the leading app vendors such as Netflix, Amazon, Xiaomi, Tencent, and Alibaba, have confirmed our findings.

\subsection{Experiments with Most-Downloaded Apps}  \label{sec:study-1}



We test data-clone attacks with $234$ popular apps from American and Chinese Android app markets,
where have the largest user base in the world. The selection criteria are: 1) the app is among the top 300 apps in that market; 2) it has more than 1 million downloads. After that, we have to install each app on a real device to test whether it can work properly. For example, some apps have regional restrictions. Finally, we obtain 114 top apps from Google Play store and 120 top apps from Huawei/Xiaomi app markets. Their distributions are shown in the second column of Table~\ref{table:three-environments}.
One of them is the subscription-based app, which relies on the user's regular payments to provide a better service. Besides, it also uses the number of connected clients in their pricing model. The citation~\cite{subscription-app} provides more details to advocate the subscription-based app economy.
For the smart home apps, we also purchase related smart home devices, including
one smart lock, two security cameras, and one smart light bulb, to test whether we can control them after launching a data-clone attack.

There are three ways to collect backup data from victim users for launching a data-clone attack. (1) Like the assumption held by related work ~\cite{Junsung18,Suwan16}, attackers either have physical access to the victim's rooted device or the malware to steal credential data has been installed on the rooted device. (2) If we assume that the users' devices are not rooted, attackers can still exploit phone-clone app vulnerabilities to intercept private user data. CVE-2019-15843 is such a zero-day vulnerability we found. We exploit this vulnerability and perform the MitM attack during data transmission. For example, when the app sets up a Wi-Fi hotspot to transfer data between two phones, we can perform ARP spoofing to successfully intercept data frames on the WLAN and then launch a data-clone attack. A concurrent work from ACSAC'20~\cite{SiqiL20}, demonstrates this type of vulnerability is popular. (3) As the users who perform the subscription fraud has full control of the device, they can use the OEM-made phone clone app to test subscription apps.

The third column of Table~\ref{table:three-environments} lists the number of successes when performing data-clone attacks with real devices.
We can automatically log in to $131$ out of $234$ apps. Table~\ref{table:real-success} shows the examples of these compromised apps, including
prominent apps that have been downloaded for more than one billion times (e.g., Facebook, WhatsApp, QQ, and Sina Weibo).
The attacks on smart home apps result in a more severe consequence, because we can remotely control all smart devices associated with our tested smart home apps.
For example, we are able to unlock the smart lock and turn off the smart light bulb and security cameras.

\subsection{Device-Consistency Check}

For the remaining $103$ failed cases, when we run them in the new device, they exhibit one of the following responses:
1) the app terminates and exits; 2) the app requests the user to type ID and password again. Many apps also pop up a new window
showing that the app is running on a different device. Therefore, it is very likely that these apps have already detected the change of device and
thus disabled the automatic login. To confirm our conjecture, we clone these apps to an Xposed-based
device-attribute editing tool, XxsqManager~\cite{XX}. It provides a virtual environment on top of the Android framework,
in which a user can edit device attributes via API hooking and thus deceive guest apps.

\begin{table}
	\centering
	\caption{The number of successes when performing data-clone attacks with real devices, Xposed-based sandbox, and \mytool, respectively. Xposed-based sandbox and \mytool\ have been configured to match the victim phone's profiles.}
	\label{table:three-environments}
	\vspace{-1mm}
	\resizebox{0.49\textwidth}{!}{
        \begin{tabular}{l r r r r}
		            	             &  \multicolumn{1}{c}{\#Apps}    & \multicolumn{1}{c}{Real device}    & \multicolumn{1}{c}{Xposed}   &  \multicolumn{1}{c}{\mytool}  \\
		    \toprule
            Social media             &  104       & 65               & 73              &  104          \\
			Payment          &  99       & 39               & 46              &  99         \\
			Subscription             &  29        & 25               & 27              &  29          \\
            Smart home               &  2         & 2                & 2               &  2            \\  \midrule
            Sum                      &  234       & 131              & 148             &  234          \\
			\bottomrule
		\end{tabular}	
	    }
	\vspace{-5mm}
\end{table}

We install XxsqManager in Huawei Honor 8.
This tool provides $65$ configuration options, and we edit all of them as the same profiles
with our old phone (Xiaomi Redmi Note 4). For the attack model of paying-subscriber fraud,
as fraudsters own the device in advance, they can run a third-party device information tool to
collect complete device artifact data.
The ``Xposed'' column of Table~\ref{table:three-environments} shows the number of successes when performing data-clone attacks with
an Xposed-based sandbox. Compared to the experiment with real devices, we have 17 compromised apps that are newly added (see Table~\ref{table:xpose-success}).
Our new experiment confirms that some apps have already
performed some device-consistency checks to secure their auto-login functions,
but their detections can be easily evaded by the user-level API-hooking mechanism.

\begin{table} [t]
  \caption{The successful examples of data-clone attacks with real devices. Due to the space limit, we only list top apps in the number of downloads.}
  \vspace{-1mm}
  \centering
  \label{table:real-success}
  \resizebox{0.49\textwidth}{!}{
  \begin{tabular}{cc}
    {Type} & {Apps}\\
    \toprule
     \multirow{3}{*}{Social media}     &  Facebook, QQ, Instagram, Snapchat, \\
                                       &  WhatsApp, Messenger, Tinder, Telegram, \\
                                       &  Pinterest, Sina Weibo  \\
     \midrule
     \multirow{3}{*}{Payment}  &  Pinduoduo, DiDi, Letgo, iHerb, \\
                                       &  OfferUp, Postmark, Shpock,  \\
                                       &  GoFundMe, Banggood, Lazada \\
     \midrule
     \multirow{3}{*}{Subscription}     &  Netflix, Amazon Prime Video, KKBox, Hulu, \\
                                       &  BBC News, Youku Video, Amazon Music, \\
                                       &  iQiyi, Netease Cloud Music \\
     \midrule
     Smart home                        &  Mi Home, 360 Smart Camera          \\

  \bottomrule
\end{tabular}}
\end{table}

\begin{table}
  \caption{The successful cases that are newly added when performing data-clone attacks in Xposed-based sandbox.}
  \vspace{-1mm}
  \centering
  \label{table:xpose-success}
  \resizebox{0.49\textwidth}{!}{
  \begin{tabular}{cc}
    {Type} & {Apps}\\
    \toprule
     \multirow{2}{*}{Social media}     &   LINE, Microsoft Outlook, Douban, Toutiao, \\
                                       &   Baidu Tieba, TikTok, Douyin, Wickr, BIGO LIVE\\
     \midrule
     \multirow{2}{*}{Payment}  &  Best Buy, NetEase Kaola,  \\
                                       &  Ctrip, 5miles, Geek, KFC \\
     \midrule
     {Subscription}                    &  TuneIn Radio, Qingting FM \\
  \bottomrule
\end{tabular}}
\vspace{-5mm}
\end{table}

However,  the app-virtualization technique is not entirely transparent to guest apps~\cite{DualInstance}. For example,
Xposed's hooking mechanism leaves identifiable fingerprints in package names, call stack methods, suspicious native methods, and
shared objects loaded into memory~\cite{XposedChecker}.
We find some cloned apps such as Alipay and Apple Music can also detect the existence of Xposed-based sandbox and
thus prevent data-clone attacks. In what follows, we explore OS virtualization to provide a transparent and customizable virtual environment.


\section{{\mytool} System Design} \label{sec:system-design}

\begin{figure*} [t]
\centering
\includegraphics[width=0.88\textwidth]{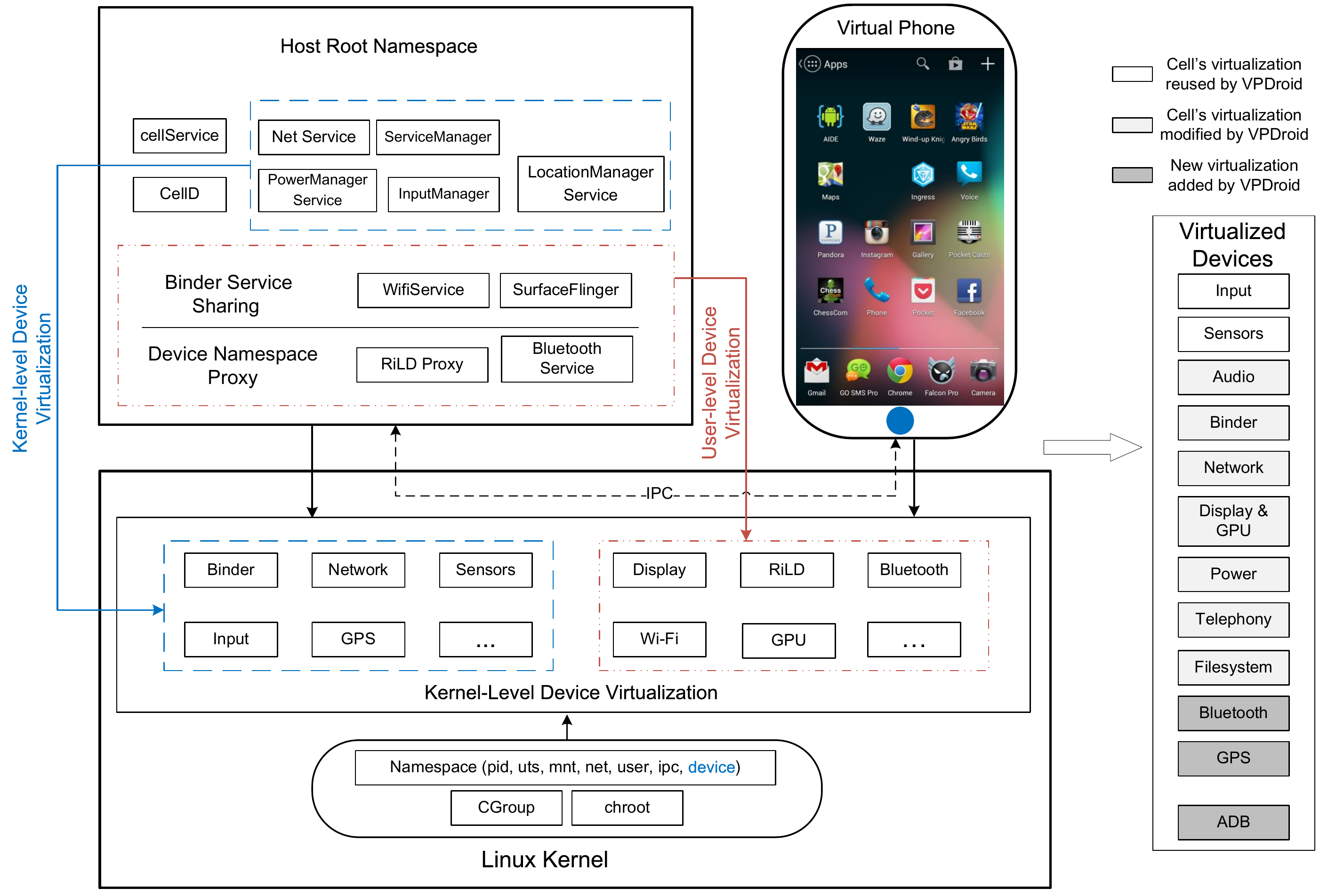}
\caption{VPDroid's device virtualization architecture (kernel-level \& user-level) and our changes to Cells.}
\label{fig:VPDroid}
\vspace{-6mm}
\end{figure*}

We develop a lightweight Android OS-level virtualization architecture, \emph{\mytool},
to assist apps' account security testing. With \mytool, security analysts are able to configure different
device attributes according to a target phone's profiles and then boot up
a virtual phone (VP) environment that closely approximates the target device.
Moreover, our solution enables device-attribute editing operations not to interfere with
the host device's normal operations. To deceive the cloned apps into thinking the smartphone
is not changed, \mytool\ has to meet two requirements (\textbf{Req1} \& \textbf{Req2}):

\begin{enumerate}
  \item \textbf{Req1}: the VP always gets direct access to hardware devices; this design provides a close-to-native virtual environment with high performance.
  \item \textbf{Req2}: user-mode apps in the VP are imperceptible to the change of device; this requires our virtualization and device-attribute customization functions are invisible to user-mode apps running in the VP.
  \end{enumerate}


\mytool\ is built on top of Cells~\cite{Cells}, because its foreground VP design meets \textbf{Req1}.
However, Cells exhibits three major limitations. 1) Cells fails to meet \textbf{Req2}: it is not designed to edit device attributes. 2) Like API-hooking, Cells's user-level device virtualization
modifies the VP's application framework layer, which can be detected by VP's apps.
3) Cells's kernel-level device virtualization to many hardware devices are not compatible with Android 6.0 and later versions anymore.
We improve Cells significantly to achieve our requirements on mainstream Android versions.


\subsection{Overview}

Figure~\ref{fig:VPDroid} provides an overview of \mytool's system  architecture.
Please note that, as a virtualization framework, VPDroid can smoothly run five virtual phones.
However, only the VP running in the foreground can always directly access all hardware devices,
which is indispensable to satisfying both \textbf{Req1} and \textbf{Req2}. Therefore,
we maintain one VP in this paper. The isolated VP runs a stock Android userspace environment. \mytool\
utilizes Linux namespaces as well as the \texttt{device} namespace introduced by Cells to transparently remap OS resource identifiers to
the VP. The VP has its private \texttt{device} namespace so that it does not interfere with the host.

We keep Cells's kernel-level device virtualization methods that still work in recent Android versions,
including Input (e.g., touchscreen and input buttons ) and Sensors (e.g, accelerometer and light sensors).
We also keep the custom process, ``CellD'', in the host device's root namespace. CellD manages the starting and switching of VPs,
and it also coordinates our ADB virtualization; ADB is used for copying data to the VP.
Since Android 8.0, Android OS has introduced a new vendor interface between the Android OS framework and the vendor implementation~\cite{Treble}.
We improve Cells's kernel-level device virtualization methods in the Binder, power management, and core network resource to be compatible with device changes in new Android
systems. Our key method is to rewrite the source code of kernel drivers so that they are aware of the \texttt{device}
namespace. Besides, we add GPS virtualization by rewriting ``/dev/gss'' driver to support multiple
connections.

\mytool\ system development is heavy in engineering.
In the next two sections, we present our two significant improvements to Cells.
First, we design a new user-level device virtualization solution with better portability and
transparency than Cells (see $\S$\ref{sec:new-user-level}). Second, \mytool\ can customize the VP's device attributes, but this function is not offered by Cells (see $\S$\ref{sec:device-customization}).

\section{New User-level Device Virtualization}  \label{sec:new-user-level}

For Cells's obsolete device virtualization solutions, rewriting every kernel driver is error-prone and complicated.
Especially, some hardware vendors provide proprietary software stacks that are completely closed source.
Without hardware vendor's support, it would be difficult, if not impossible, to virtualize them in the kernel.
\mytool's user-level virtualization offers a flexible and portable alternative \emph{without leaving any in-guest
virtualization component}.
Our mechanism contains two methods to virtualize different devices.

\textbf{Binder Service Sharing.} For the system services that are registered in ServiceManager (e.g., WifiService),
we develop a new way to virtualize them. Binder is the inter-process communication (IPC) mechanism in Android. The Binder driver is a custom pseudo
driver with no corresponding physical device.
We first modify the Binder-driver data structure (e.g., \texttt{context\_mgr\_node}, \texttt{procs}, and \texttt{dead\_nodes}) to ensure that
the VP has its own Binder-driver data structure.
In addition, we create a new specific handler in Binder's data structure and make it point to the host's \texttt{context\_mgr\_node}.
As \texttt{context\_mgr\_node} is associated with ServiceManager, with this handler, the VP can access
the host phone's ServiceManager node. Therefore, this mechanism allows a service process in the VP to share the corresponding service in the host system.
Furthermore, we leverage SELinux technology to enforce which services in the host system can be shared by the VP.
In \mytool, WiFi configuration are virtualized in this style.

\textbf{Device Namespace Proxy.} For the anonymous services that are not registered in ServiceManager, as the kernel does not have their \texttt{binder\_node} and \texttt{binder\_ref} structures,
we cannot apply binder service sharing. Instead, we virtualize them by creating a new device namespace proxy in the host userspace only.
This proxy communicates the VP service through Binder service sharing or socket. It distinguishes the VP's request from the host's request by
their associated \texttt{device} namespaces and interacts with kernel drivers to respond to the VP's request.
In \mytool, telephone are virtualized using this method.

Next, we use WiFi configuration and telephone as examples to present our
new user-level device virtualization mechanism.

\subsection {Binder Service Sharing:  WiFi Configuration} \label{sec:our-updates}

\begin{figure} [!t]
\centering
\includegraphics[width=0.5\textwidth]{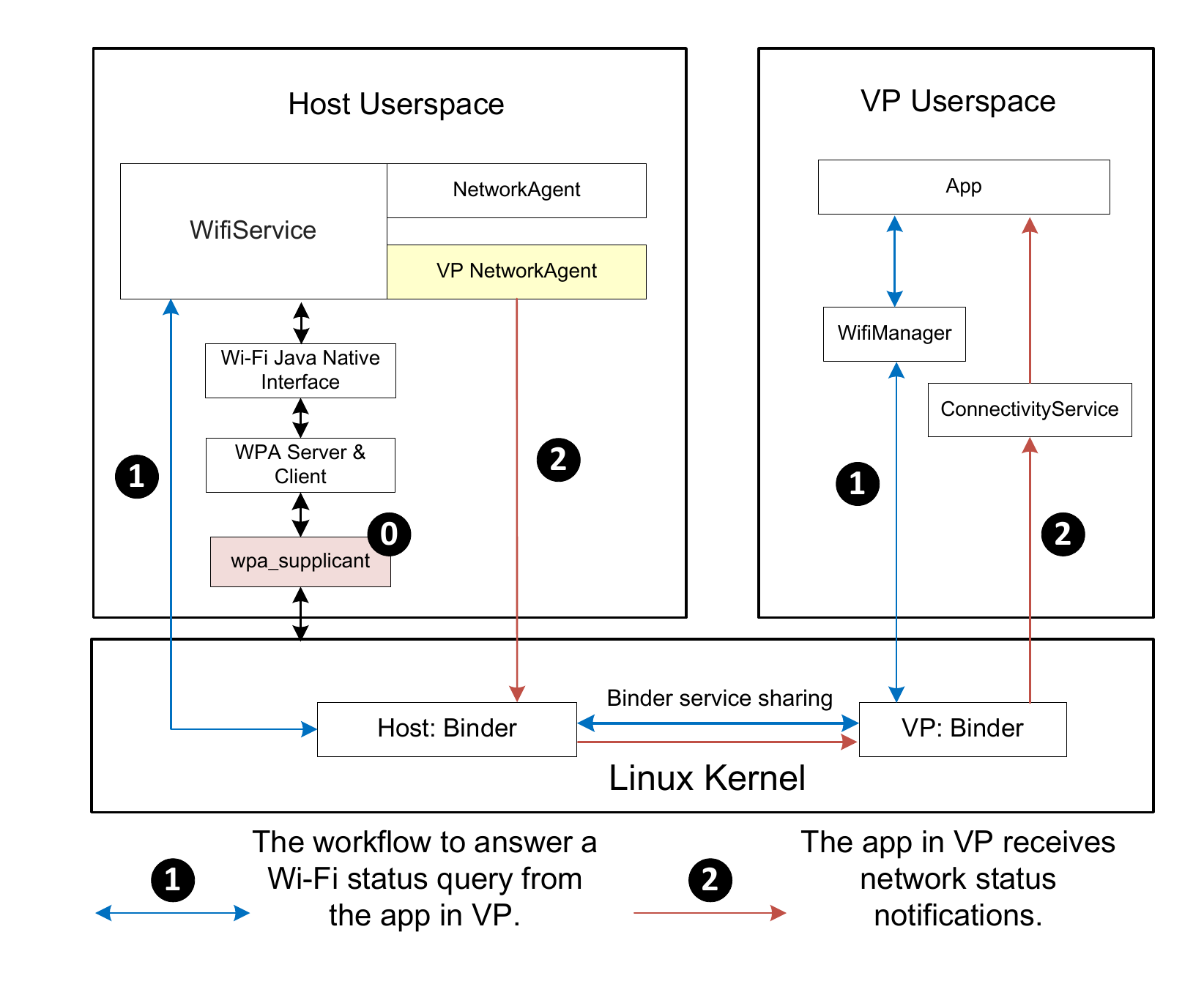}
\caption{\mytool\ leverages binder service sharing mechanism to virtualize wireless configuration management.}
\label{fig:wifi-virtual}
\vspace{-6mm}
\end{figure}

WiFi configuration and status notifications occur in the userspace. ``wpa\_supplicant''
is a user-level library that contains wireless network service code (\circled{0} in Figure~\ref{fig:wifi-virtual}).
Cells replaces ``wpa\_supplicant'' inside the VP with a WiFi proxy, which forwards all configuration requests from the VP
to the host's ``wpa\_supplicant''. In contrast, we leverage our proposed Binder service sharing to achieve the same goal,
but leaving no change in the VP's userspace.
In the Android system, WifiService calls the library of ``wpa\_supplicant'' to detect WiFi connections, and
such information is sent through NetworkAgent to ConnectivityService, which answers app queries about the state of network connectivity.

We use the Binder service sharing mechanism to share WifiService between the VP and the host system. The blue two-way line in Figure~\ref{fig:wifi-virtual}
represents the  workflow to answer a WiFi status query from the VP's app (\circled{1}).
Besides, we create a new NetworkAgent in the host system and bind it to the VP's \texttt{device} namespace.
As shown in Figure~\ref{fig:wifi-virtual}'s red line (\circled{2}), to automatically forward network status notifications to the VP,
we also use the Binder service sharing mechanism to transfer the new NetworkAgent to the VP's ConnectivityService.
Finally, the VP succeeds in receiving the status notifications of WiFi connectivity.

\subsection{Device Namespace Proxy: Telephony} \label{sec:telephony}

\begin{figure} [t]
\centering
\includegraphics[width=0.43\textwidth]{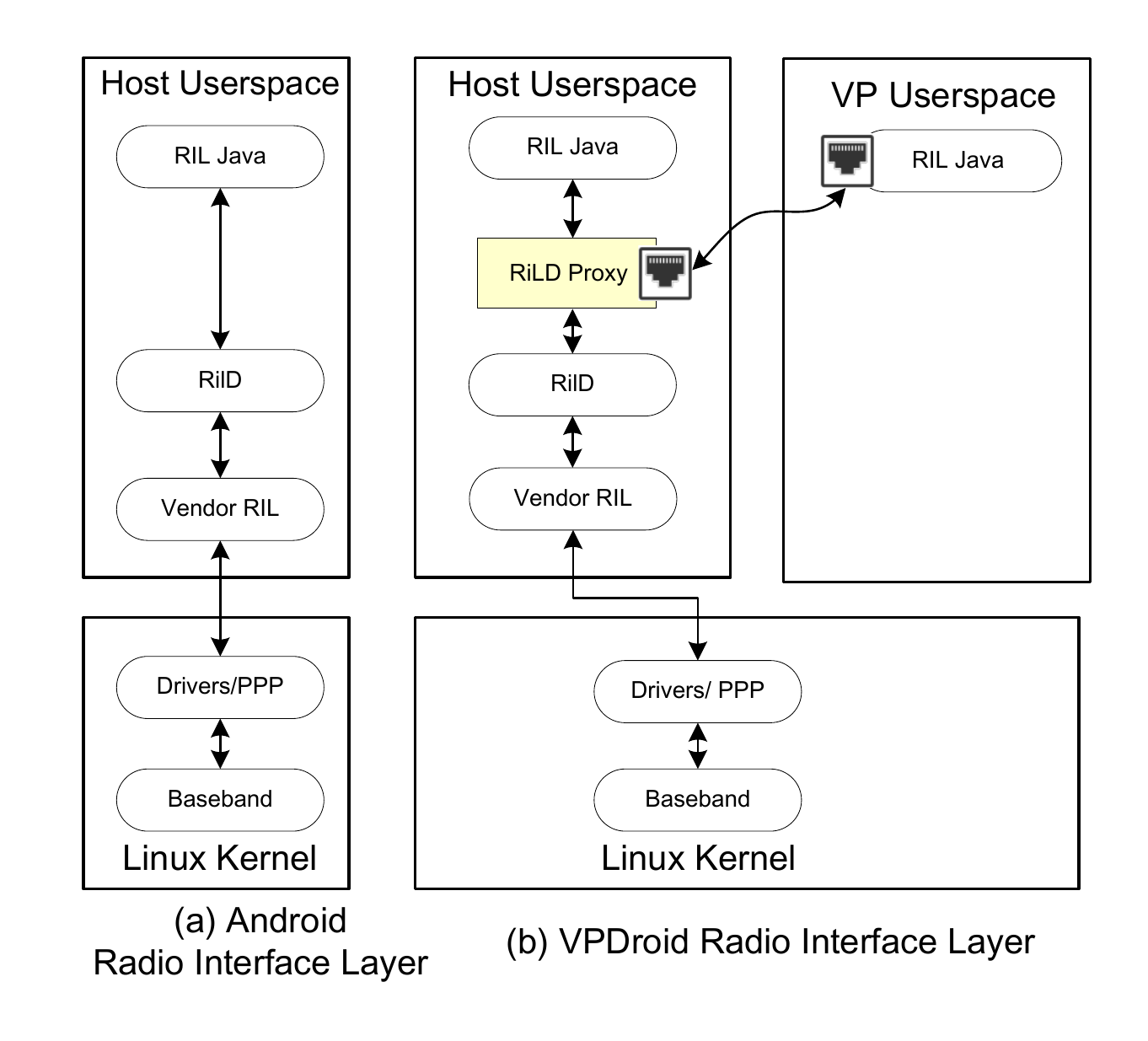}
\vspace{-2mm}
\caption{\mytool\ virtualizes telephony by creating a device namespace proxy in the host userspace only.}
\label{fig:telephony-virtual}
\vspace{-6mm}
\end{figure}

Figure~\ref{fig:telephony-virtual} (a) shows the standard Android
Radio Interface Layer. As smartphone vendors customize their own proprietary radio stack,
Cells adopts a user-level device namespace proxy to provide a separate telephony functionality for a VP.
A VP has its own proxy Radio Interface Layer (RIL) library.
The RIL proxy is loaded by Radio Interface Layer Daemon (RilD) and
connects to CellD running in the host's root namespace,
and CellD, in turn, communicates the hardware vendor library
to respond to the VP's requests. However, the RIL proxy is visible to VP's apps, which does not meet our \textbf{Req2}.

As shown in Figure~\ref{fig:telephony-virtual}(b), we implement a
socket-interface based proxy scheme only in the host userspace, and it does not require the assistance of CellD.
In the host's Radio Interface Layer, we create a RiLD proxy between the communication flow of Android telephony Java libraries (RIL Java) and RilD.
Then we create another two standard Unix Domain sockets in the proxy. One socket connects to the RIL Java of the VP, and the other one connects to the RIL Java of the host system.
The RIL Java in the VP communicates with the host system's proxy, and the proxy passes the communication data (e.g., dial request and SIM) to the host system's RilD.
In turn, the RilD proxy passes the VP-related arguments (e.g., call ring and signal strength) to the VP's RIL Java over a socket.

\begin{figure*} [!t]
\centering
\includegraphics[width=1.0\textwidth]{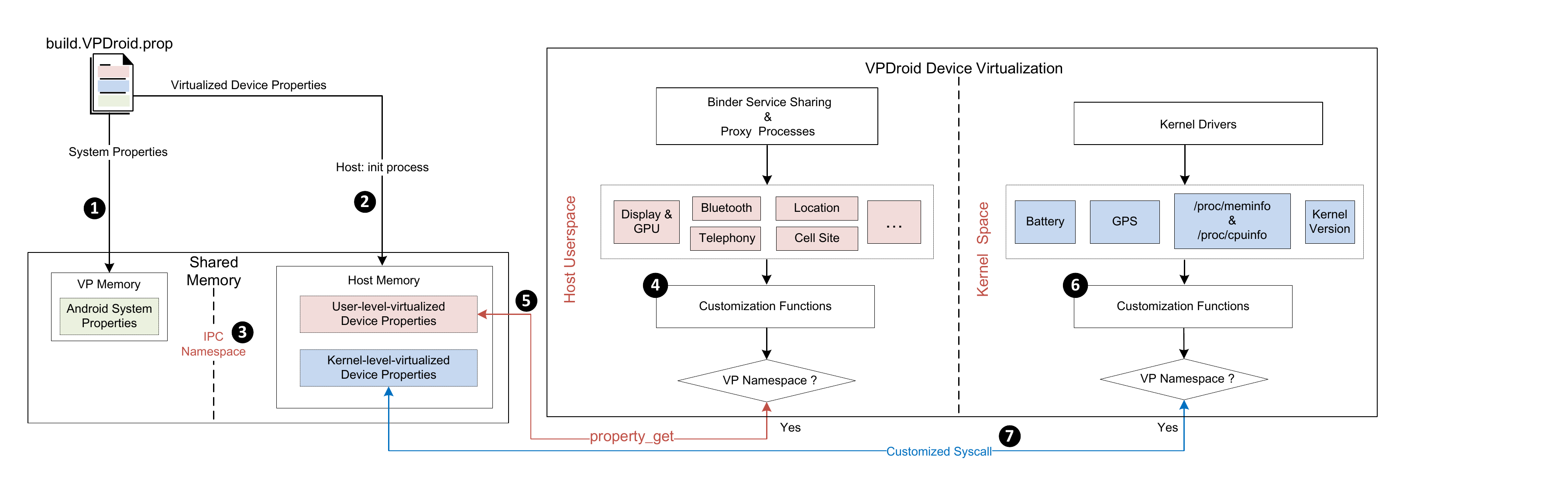}
\caption{\mytool's workflow of customizing device-specific attributes.}
\label{fig:deep-isolation}
\vspace{-1mm}
\end{figure*}

\section {Customize the VP's Device Attributes}  \label{sec:device-customization}

Based on the new Android OS-level virtualization framework, we go one step further to customize the VP's device attributes.
Figure~\ref{fig:deep-isolation} shows the workflow. VPDroid users provide a configuration file ``build.VPDroid.prop'' in advance,
which stores device-specific attributes in the form of key-value pairs.
We classify these key-value pairs into three categories: Android system properties,
user-level-virtualized device properties, and kernel-level-virtualized device properties.
Each category has a different customization method.
Besides, we incorporate multiple namespaces to isolate our customization.


\textbf{Android System Properties. }
Android system properties, stored in the init process's shared memory,
describe the configuration information of the smartphone, such as brand, model, serial number, IMEI, and Android ID.
These const values have nothing to do with our device virtualization.
Other processes enquire about Android system properties at run time by calling ``property\_get'', an API for
native code to read the data in the shared memory from other processes.
When booting up the VP, we enforce the VP's init process to load the customized Android system properties from ``build.VPDroid.prop''
into the VP's shared memory space (\circled{1} in Figure~\ref{fig:deep-isolation}).

%
%

\textbf{User-Level and  Kernel-Level Customization. }
The customized data for both user-level-virtualized and kernel-level-virtualized devices are loaded into the host init process's shared memory (\circled{2}).
We use the IPC namespace for the host and VP shared memory isolation (\circled{3}).
Our customization functions are located at the places where we just finish user-level device virtualization (\circled{4})
or kernel-level device virtualization (\circled{6}).
All of the customization functions work in a similar style.
They first determine whether the current query request is from the VP or the host by checking the associated \texttt{device} namespace.
For a user-level customization function, if the query is from the VP, it calls ``property\_get'' to get the customized data from the host's shared memory (\circled{5})
and then returns the custom data to the VP.
However, for a kernel-level customization function, the customized data loaded into the init process have no privilege to enter the kernel space.
Therefore, we create a new system call to copy data from the userspace to the kernel space (\circled{7}).

 \begin{figure*} [t]
  \centering
  \subfigure[Normalized Nexus 6p-1 results]{
    \label{subfig:a} 
    \includegraphics[width=0.31\textwidth]{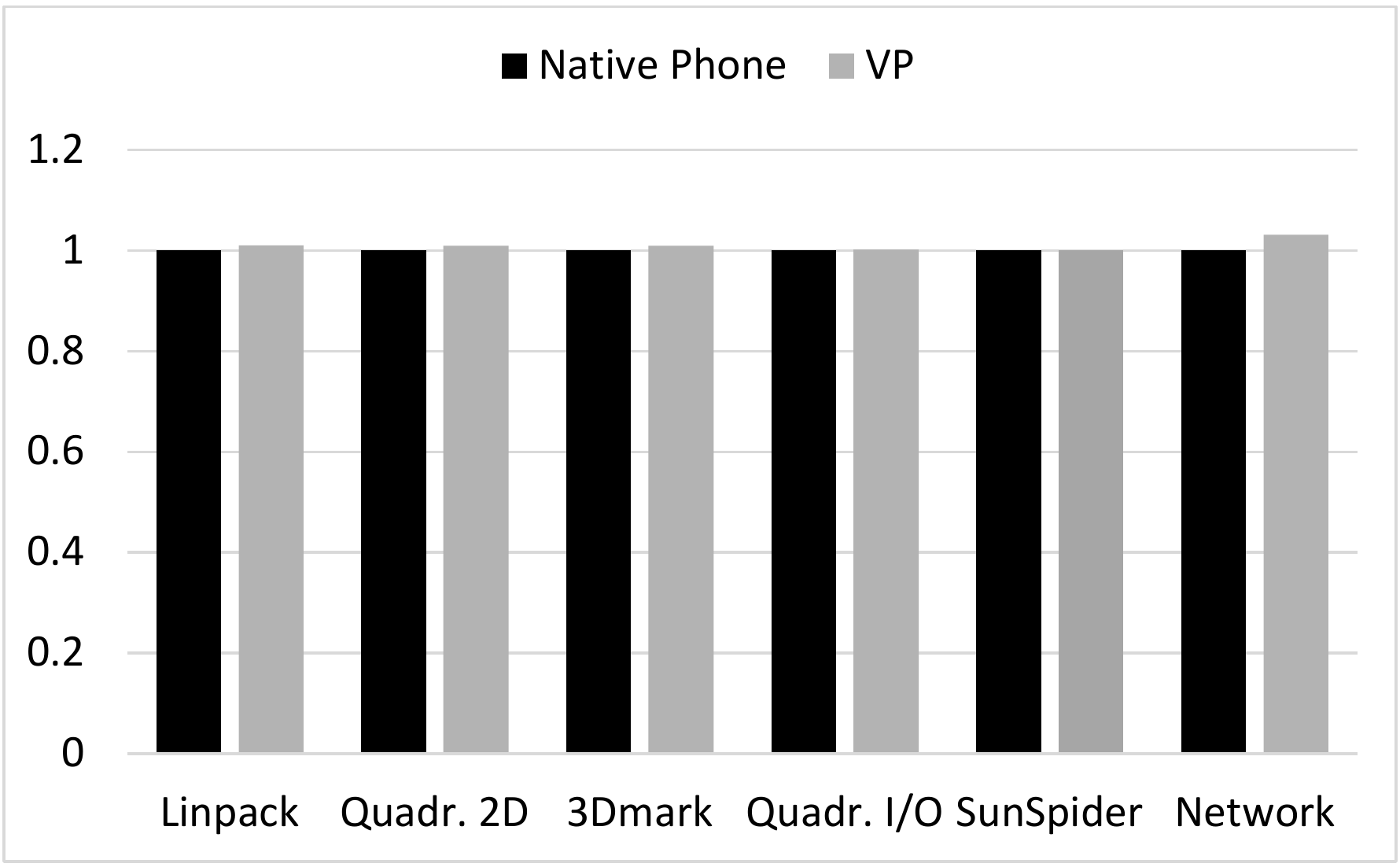}}
    \subfigure[Normalized Nexus 6p-2 results]{
    \label{subfig:b} 
    \includegraphics[width=0.32\textwidth]{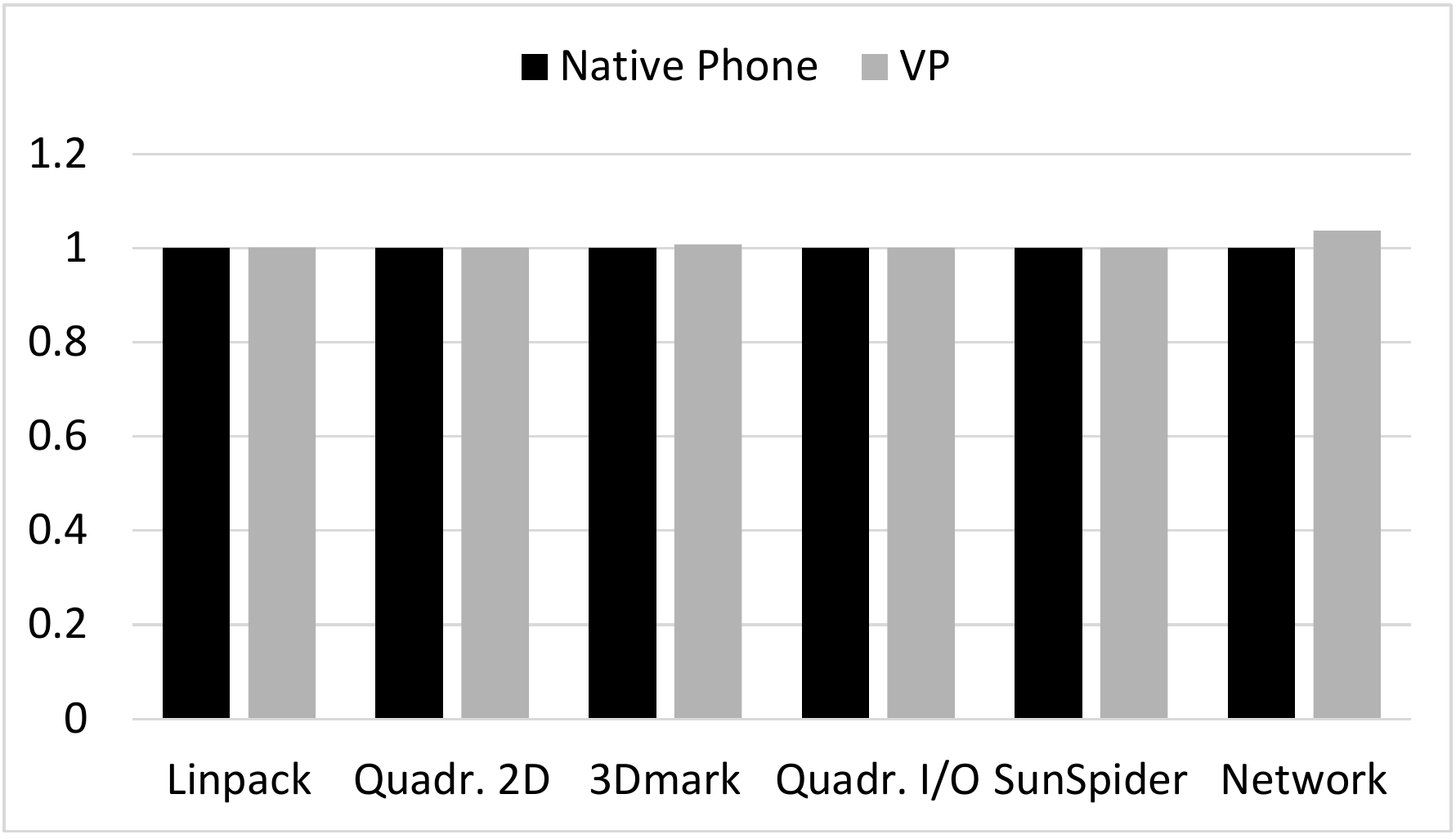}}
  \subfigure[Normalized Nexus 6p-1 + music results]{
    \label{subfig:c} 
    \includegraphics[width=0.32\textwidth]{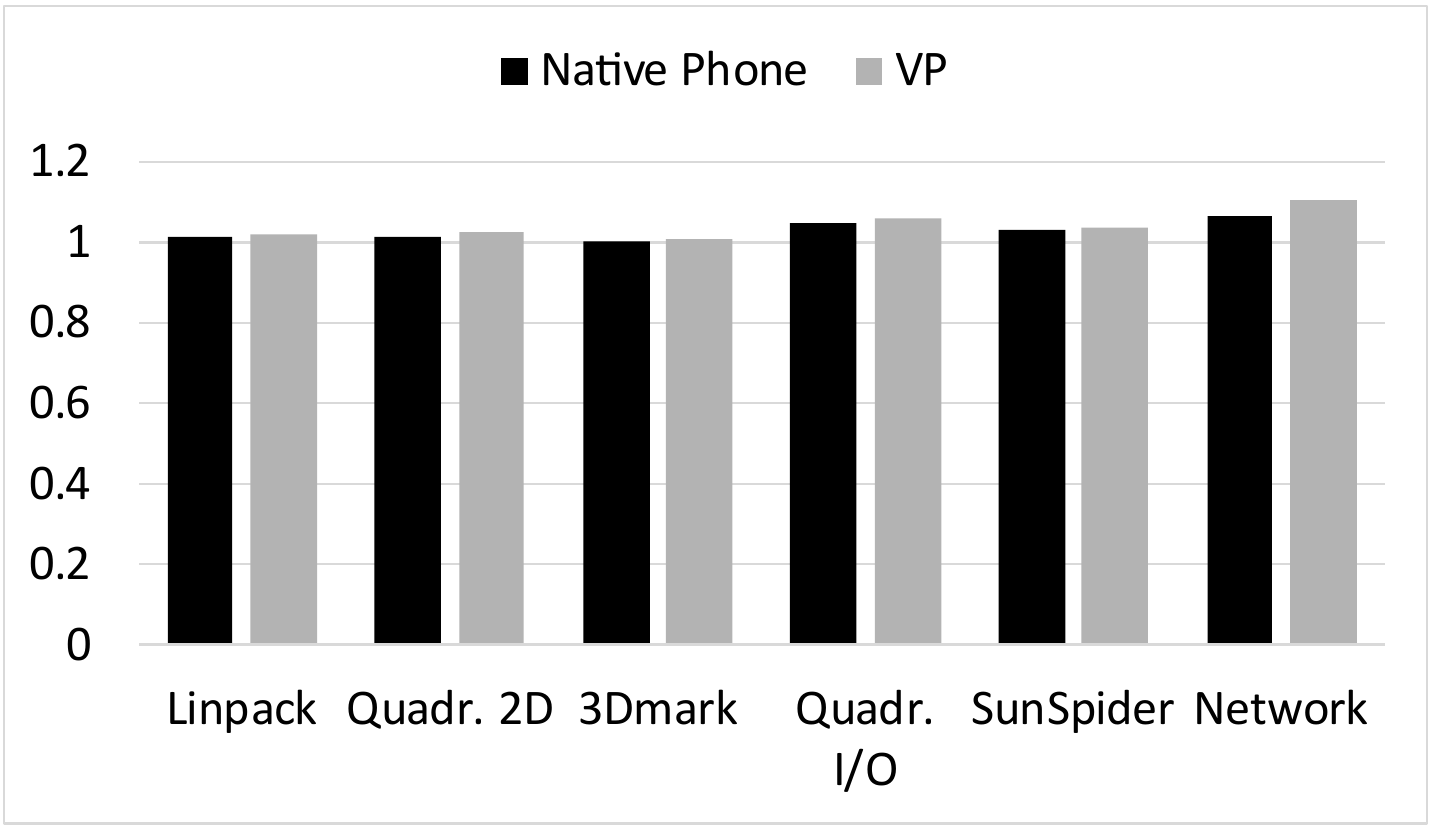}} \\ \vspace{-2mm}
  \subfigure[Normalized Nexus 6p-2 + music results]{
    \label{subfig:d} 
    \includegraphics[width=0.33\textwidth]{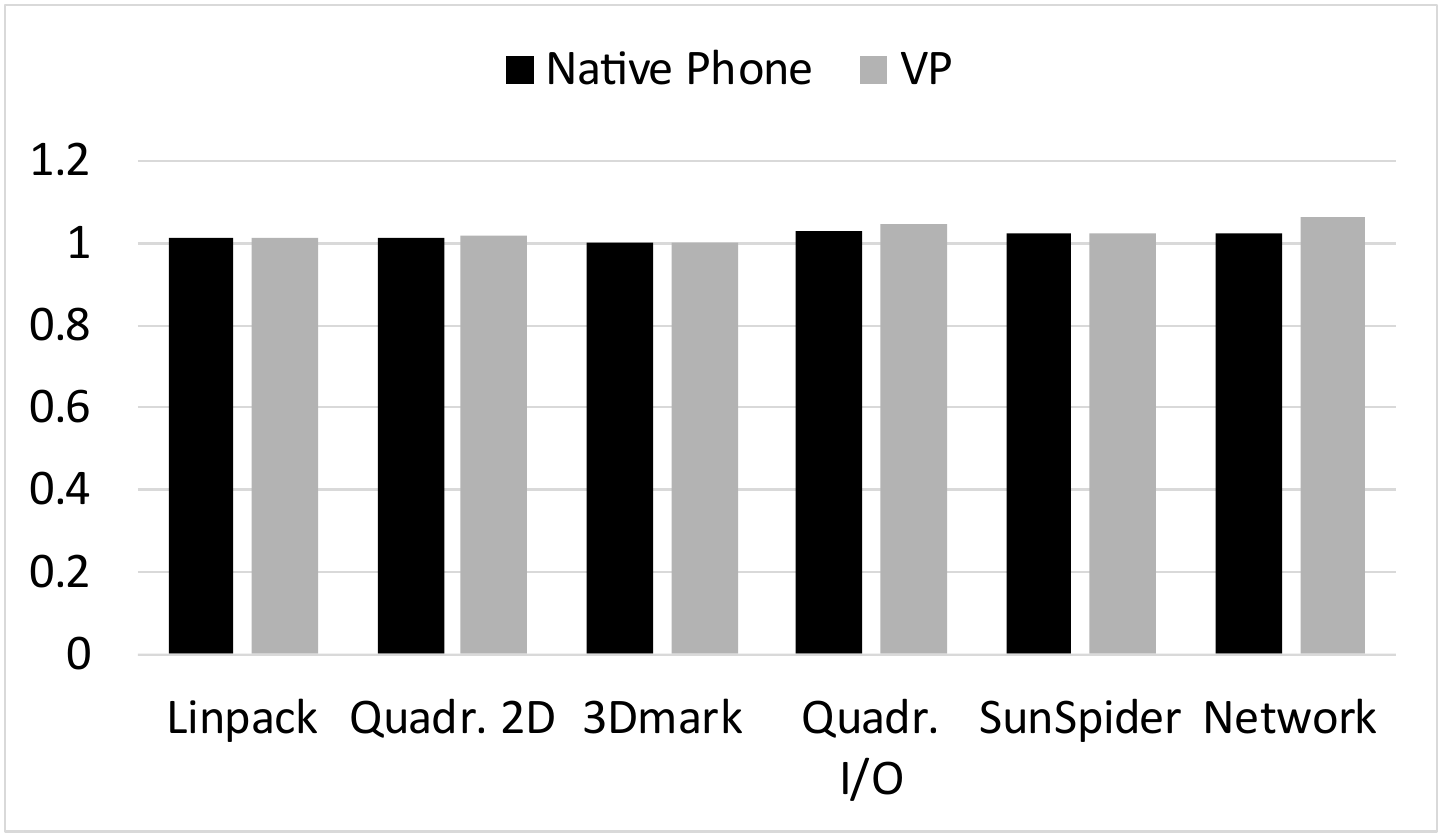}}
    \subfigure[Nexus 6p-1 memory usage in MB]{
    \label{subfig:e} 
    \includegraphics[width=0.32\textwidth]{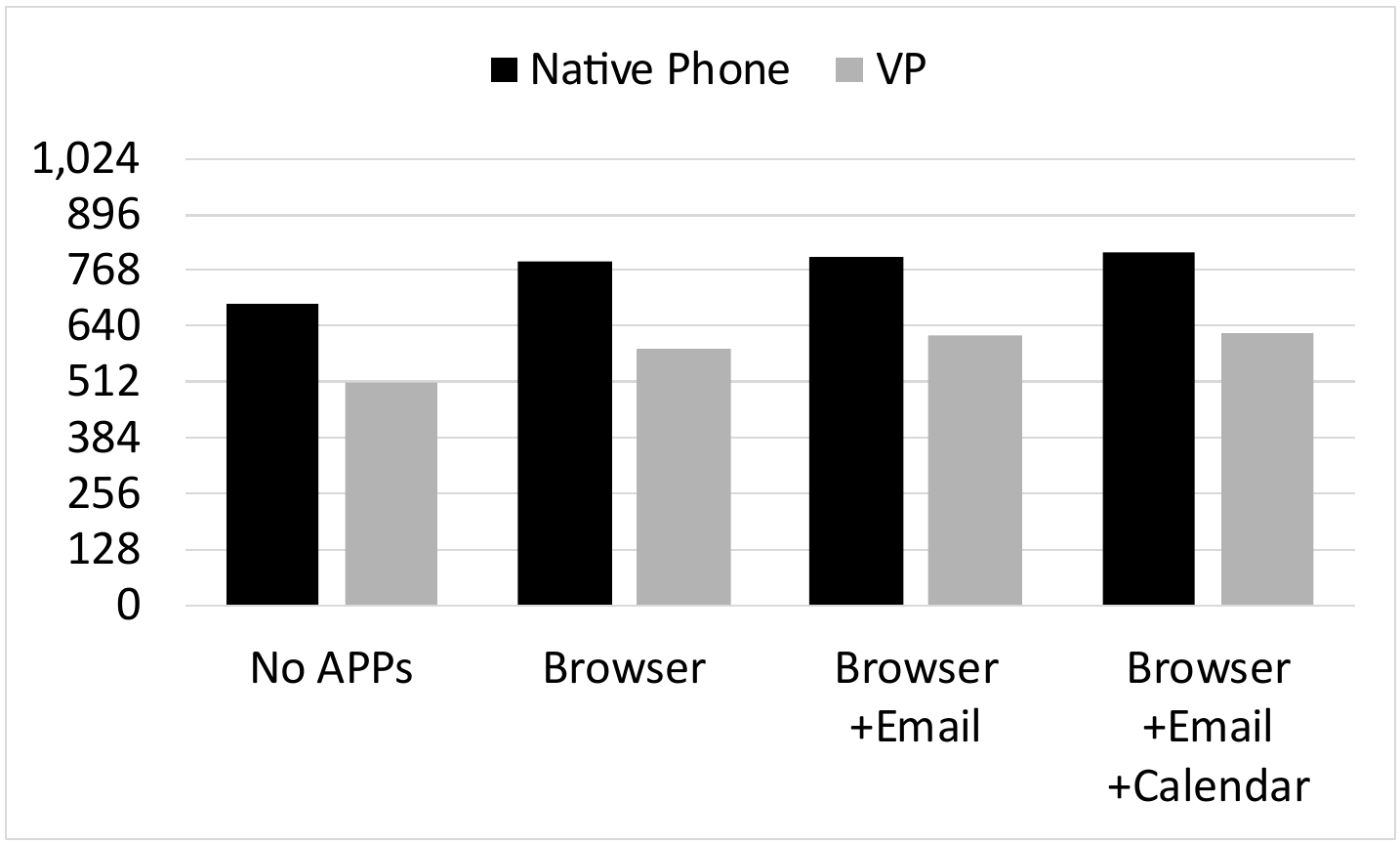}}
  \subfigure[Nexus 6p-2 memory usage in MB]{
    \label{subfig:f} 
    \includegraphics[width=0.315\textwidth]{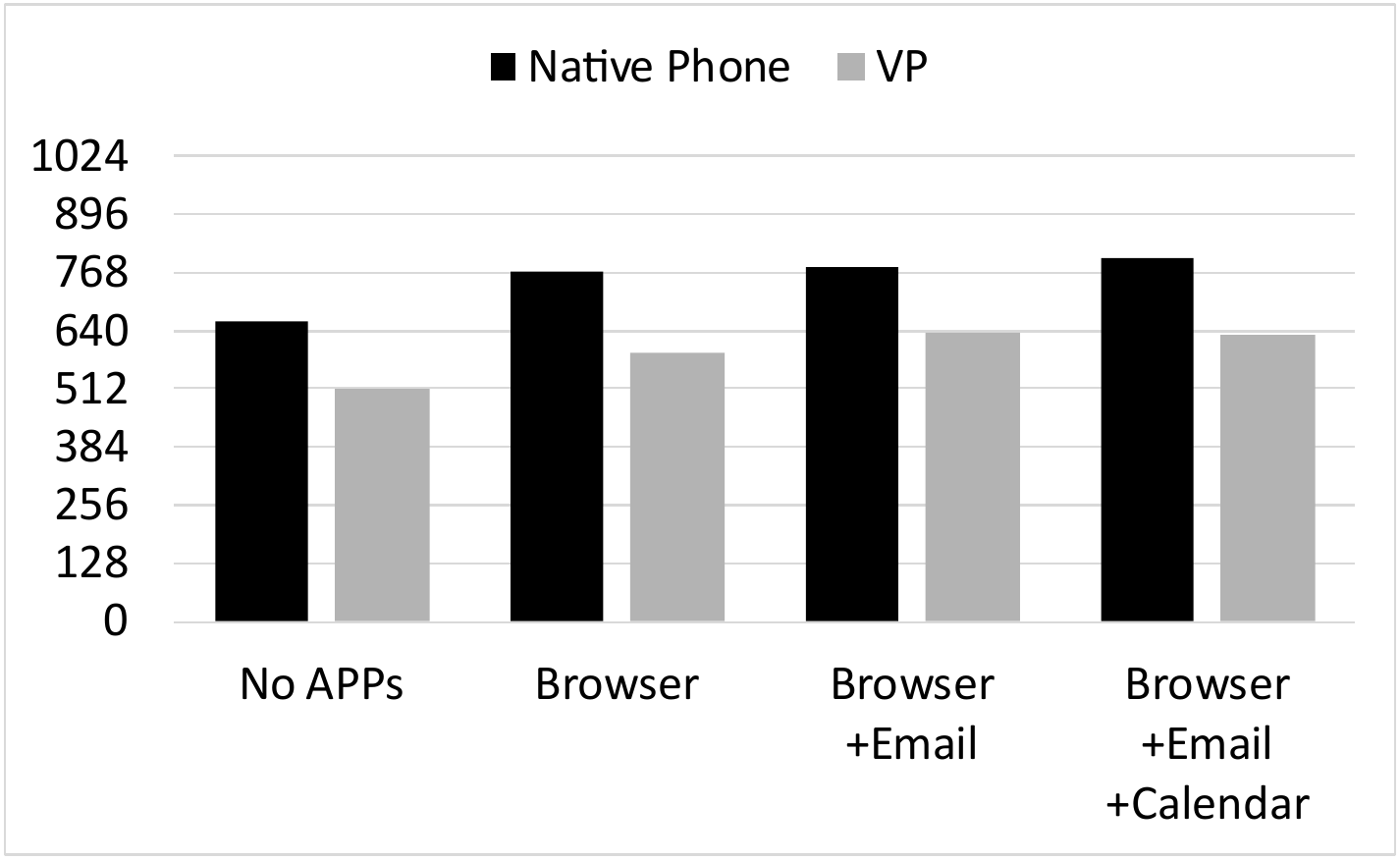}}
  \caption{\mytool's performance measurements. ``Quadr.'' is short for ``Quadrant''}
  \label{fig:performance} 
  \vspace{-5mm}
\end{figure*}

\textbf{The Advantages of \mytool\ Customization. }
Compared with existing Android device-attribute editing tools~\cite{Bianchi17,XX}, our customization solution revels distinct advantages.
\textbf{First}, all of our customization functions do not rely on any user-level API hooking mechanism, and they are executed outside of
the VP's runtime environment. This means our device customization is invisible to VP's user-mode apps.
Although our user-level device virtualization allows the VP's process to share certain services in the host system,
with the \texttt{device} namespace isolation, a user-mode app running in the VP  is still unaware of device-specific differences.
\textbf{Second}, our VP's customization does not interfere with normal operations on the host device.
System modifications without leveraging OS-level virtualization lack flexibility and compatibility.
Besides, they are very difficult to achieve the same transparent and stealthy capability as \mytool,
as blindly changing return values of APIs/syscalls is likely to cause system crashes or exceptions (e.g., Bluetooth system services keep restarting).
Due to the multiplexing of hardware devices, VPDroid avoids incompatibility issues by
decoupling device-attribute editing operations from normal operations on the host device.

\mytool\ now can support customizing $101$ device configuration options,
which span a broad spectrum of device attributes. We collect these options from existing work on the Android device artifact detection.
To the best of our knowledge,  \mytool\ offers the most comprehensive Android device-attribute editing options so far.
%


\section{\mytool\ Evaluation}

The VP images are created on a PC and downloaded to the host device via USB.
We provide a control center app for \mytool\ users to efficiently switch between the VP and the host system.
To start a new VP to simulate a different device, a user takes the following three steps:
1) exiting the original VP; 2) updating and replacing a new ``build.VPDroid.prop'' configuration file;
3) stating a new VP via the control center app.
This section first provides performance measurements to show that \mytool\ reveals native performance.
In our second experiment, we use the data-clone attack as a case study to evaluate
\mytool's capability on device-attribute customization. Our results show that
\mytool\ substantially increases the success rate of the data-clone attack.

%


\subsection{Performance Measurements}
We measure runtime overhead and memory usage using two Google Nexus 6P phones that are different in CPU model and ROM size:
Nexus 6p-1 (ARM Cortex-A53, Adreno 430 GPU, 3G RAM, and 32G ROM), and Nexus 6p-2 (ARM Cortex-A57, Adreno 430 GPU, 3G RAM, and 64G ROM).
We follow similar experimental settings with Cells's paper in SOSP'11~\cite{Cells}. Our runtime overhead measurement contains two scenarios.
The first one is running a set of benchmark apps on \mytool's VP and a native phone, respectively.
The second one is running the same benchmark apps on the VP and the native phone, but simultaneously with an additional background music player workload.
All results are normalized against the performance of running the same benchmark apps on the latest manufacturer stock
image available for Google Nexus 6P, but without the background workload.
Each benchmark app is designed to stress some aspect of the system performance:
Linpack (v1.1) for CPU; Quadrant advanced edition (v2.1.1) for 2D graphics and file I/O;
3DMark (v2.0.4646) for 3D graphics; SunSpider (v1.0.2) for web browsing;
and networking using BusyBox wget (v1.21.1) to download a single 409M video file through
a PC's Wi-Fi hotspot.

Figure~\ref{fig:performance} shows the normalized runtime overhead and memory usage on two
Nexus 6P phones. Compared to Cells 1-VP's data~\cite{Cells}, \mytool\
reveals the same level of variability in measurement results, and it is even better than Cells
in Quadrant I/O, SunSpider, and Network results from 5\% to 9\%. Cells's performance data were obtained
using Nexus 1 and Nexus S. We admit that the hardware upgrade caused by Nexus 6P also favorably impacts our results.
The deviations between ``VP'' and ``Native Phone'' in Figure~\ref{subfig:a} $\sim$ Figure~\ref{subfig:d}
represent the additional overhead caused by \mytool's device virtualization. The negligible deviations
indicate no user-noticeable performance difference between running in \mytool\ and running natively on the phone.
The major difference from Cells is memory usage. For example, after booting up 1-VP with no apps running,
Cells's memory usage is 128 MB, but this number increases to 512 MB for \mytool.
As would be expected, the Android OS's size is also bloating.  As shown in Figure~\ref{subfig:e} and Figure~\ref{subfig:f},
the memory usage in the VP is less than the native phone in all workload cases. The reason is due to the
lightweight OS-level virtualization, the memory consumed by kernel services only occurs at the host device.

\subsection{Virtualization-Assisted Data-Clone Attack}

We repeat our data-clone attacks with most-downloaded apps (see $\S$\ref{sec:study-1}) in \mytool.
In particular, we take Google Nexus 6P-2 as the host machine and configure the VP environment
as Xiaomi Redmi Note 4, Redmi Note 4x, Huawei Honor 6x, Honor 8, and Google Nexus 6P-1, respectively.
These five VP environments represent five victim devices, and we provide five different device-attribute configuration files for \mytool\ to load.
In spite of the diversity, we achieve the same results for all cases. The last column of Table~\ref{table:three-environments} shows the success number of
data-clone attacks in \mytool: we can compromise all of the $234$ most popular apps' accounts. Compared
with the attacks on a real device, \mytool\ wins by additional $103$ apps; among them, $86$ apps can detect Xposed-based sandbox but fail to detect \mytool.
Note that our attacks failed at first for some apps that rely on Android ``AccountManager'' APIs to manage the auto-login function (e.g., Youtube, Google Play, Twitter, and Skype).
The reason is ``AccountManager'' stores auto-login depended data
under the directory of ``/data/system\_xx/'' rather than ``/data/data/[app\_name]/''. After we copy the ``/data/system\_xx/'' folder to the virtual phone in our second try-out,
our data-clone attacks succeeded.


\begin{table} [t]
  \caption{Zero-day vulnerabilities that we found. The third column shows the vulnerability ID or the vendor's confirmation time.
  The app's name in bold represents this app takes the device-consistency check to protect its auto-login function.}
  \vspace{-1mm}
  \centering
  \label{table:zero-day}
  \resizebox{0.44\textwidth}{!}{
  \begin{tabular}{ccc}
     {App} & {Vendor} & {Vul. ID or Conf. Time}  \\
    \toprule

     Netflix                 &  Netflix  & 10/12/2019 \\
     Prime Video             &  Amazon   & 11/16/2019  \\
     \textbf{Kugou Music }            &  Tencent  & VULBOX-2019-0221059  \\  
     \textbf{Tencent Video}           &  Tencent  & 8/18/2019 \\
     iQiyi                   &  iQiyi    & 8/18/2019  \\
     Youku Video             &  Alibaba  & 8/18/2019 \\
     \textbf{WPS Office }             &  Kingsoft & 10/16/2019 \\
     \textbf{Zhihu}                   &  Zhihu    & CNVD-2019-42028 \\
     \textbf{University MOOC}         &  NetEase  & 9/29/2019  \\
     \textbf{Cloud Classroom}         &  NetEase  & 9/23/2019 \\
     Unipus                  &  Unipus   & CNVD-2019-42030 \\
     \textbf{Dragonfly FM}            &  Dragonfly  & CNVD-2019-42031 \\
     LiuLiShuo               &  LAIX     & CNVD-2019-42032 \\
     CAD Quick Look          &  Glodon   & CNVD-2019-42034 \\
     VivaVideo               &  Quwei    & CNVD-2019-42033            \\
     Yizhibo                 &  Yizhibo  & CNVD-2019-42029  \\
     Smart Camera            &  Qihoo 360 & 1/20/2020 \\
     Mi Home                 &  Xiaomi   & 3/30/2019 \\
     \textbf{Mi Mover}                &  Xiaomi   & CVE-2019-15843 \\
  \bottomrule
\end{tabular}}
\vspace{-5mm}
\end{table}

To make sure no app can detect the change of device in \mytool,
we perform another comparative experiment with an OEM-made backup app. It has the privilege to call Android Backup API~\cite{android-backup} on the same OEM phones,
so it can backup and restore user private data in ``/data'' partition.
We first use this OEM-made backup app to backup our tested apps and then restore them. Next, we keep a record of the apps whose auto-login functions still work
after backup-restore. We treat the effect of this experiment as launching a data-clone attack on the same device. However, if one of these apps fails to automatically
log in after we clone its auto-login depended data to the VP, it means this app finds \mytool's environment is different from the original device.
However, we did not find such a counterexample.
This confirms that our device virtualization and
customization are transparent to cloned apps.

\textbf{Vendor Reaction}.
Table~\ref{table:zero-day} lists the zero-day vulnerabilities we have identified.
We find that Chinese vendors take our findings more seriously than American vendors. For example, Netflix confirms our vulnerability finding, but they treat it as
a ``Single-User Fraud'' threat. We confirm that with the latest Netflix version (7.68.4), our data-clone attack without \mytool\ still succeeds.
By contrast, Chinese vendors such as Alibaba, Tencent, and iQiyi have labeled our findings as high/middle-severity vulnerabilities.
We speculate that Chinese vendors are more vulnerable to paying-subscriber fraud.
For example, iQiyi, an online video app with more than 100 million users, has labeled our report as a high-severity vulnerability and
added device-consistency checks in the new release version.
However, we have evaluated the latest version of iQiyi in \mytool\ and found that,
VPDroid can still bypass the newly added device-consistency checks.

\section{Discussion} \label{sec:discussion}

The most fundamental method against data-clone attacks is that
a mobile app never stores user credential data in local files.
However, this strategy, at the cost of sacrificing usability,
only works for critical apps that do not require frequent user interactions.
Another direction is to leverage a Trusted Execution Environment (e.g., ARM TrustZone) to encrypt/decrypt
user credential data before use. As the decryption key is stored in the TrustZone environment,
data-clone attacks cannot copy the decryption key to another device together with encrypted user credential data,
and therefore the server will fail to verify the login credentials.
The recent papers, TruApp~\cite{TruApp} and IM-Visor~\cite{IM-Visor}
explore the feasibility of protecting app integrity and sensitive data with TrustZone,
and Rubinov et al.'s work partitions a critical app automatically for TrustZone~\cite{Rubinov16}.

%

A natural response to breaking through the login-device number limit
is to monitor concurrent sessions at the app server side.
Unfortunately, the variable nature of mobile devices (e.g., the switch of WiFi hotspot and cellular data)
makes it difficult to determine an adequate number of concurrent sessions.
The previous work~\cite{Jongwon15} has pointed out that, although many apps do not permit duplicate
logins from different devices, they do allow multiple session requests from the same device ID.
Our evaluation also confirms that most apps allow maintaining two or more connections per user.
As the login from \mytool\ shows a different IP address from the victim's IP,
a possible countermeasure is to detect multiple concurrent IPs at the server side.
However, this strategy cannot completely thwart data-clone apps. For quite a few
apps, such as Facebook and the smart home apps we tested, they do allow multiple logins from different devices.


We do not assume that detecting the presence of VPDroid is strictly impossible, but it can prohibitively increase the cost.
If an app in the VP has the root privilege, it can find out the footprint of our user-level device virtualization. For example,
VP's telephony Java libraries do not interact with VP's RilD.
The auto-login function could check the consistency of some obscure device properties that are not covered by us,
and finding all of them is an open problem.


\section{Conclusion}

In this paper, we characterize, research, and evaluate the data-clone attack and its client-side countermeasure---device-consistency check.
Our technical contribution is to develop a transparent device-attribute customization platform
via Android OS-level virtualization.
Our evaluation with most-downloaded apps demonstrates that
the data-clone attack is an imminent threat, leading to great losses to the app economy and user privacy.
We wish our study and open-source \mytool\ help researchers redesign apps'
auto-login functions and evaluate the device-artifact detection capability.

%
%
%
%

\section*{Acknowledgments}
We sincerely thank ICSE 2021 anonymous reviewers for their insightful and helpful comments.
This research was supported in part by the National Natural Science Foundation of China (U1636107, 61972297) and
the National Science Foundation (NSF) under grant CNS-1850434.

{
\balance
\bibliographystyle{unsrt}
\bibliography{bib/data-clone,bib/jiang}
}


\end{document}